\documentclass[a4paper]{article}\usepackage[]{graphicx}\usepackage[]{xcolor}
\makeatletter
\def\maxwidth{ %
  \ifdim\Gin@nat@width>\linewidth
    \linewidth
  \else
    \Gin@nat@width
  \fi
}
\makeatother

\definecolor{fgcolor}{rgb}{0.345, 0.345, 0.345}

\usepackage{framed}
\makeatletter
 {\par\unskip\endMakeFramed%
 \at@end@of@kframe}
\makeatother

\definecolor{shadecolor}{rgb}{.97, .97, .97}
\definecolor{messagecolor}{rgb}{0, 0, 0}
\definecolor{warningcolor}{rgb}{1, 0, 1}
\definecolor{errorcolor}{rgb}{1, 0, 0}

\usepackage{alltt}
\usepackage{style}
\usepackage{authblk}

\usepackage[color=blue!20!white,textsize=scriptsize,textwidth=28mm]{todonotes}

\newcommand{\Ga}{\mathsf{Gamma}}

\title{A joint Bayesian framework for missing data and measurement error using integrated nested Laplace approximations}
\author[1]{Emma Skarstein}
\author[1]{Sara Martino}
\author[1,2]{Stefanie Muff}
\affil[1]{Department of Mathematical Sciences, Norwegian University of Science and Technology, Norway}
\affil[2]{Centre for Biodiversity Dynamics, Norwegian University of Science and Technology, Norway}
\date{\small{\today}}
\IfFileExists{upquote.sty}{\usepackage{upquote}}{}
\begin{document}



\maketitle

\begin{abstract}

Measurement error (ME) and missing values in covariates are often unavoidable in disciplines that deal with data, and both problems have separately received considerable attention during the past decades. However, while most researchers are familiar with methods for treating missing data, accounting for ME in covariates of regression models is less common. In addition, ME and missing data are typically treated as two separate problems, despite practical and theoretical similarities. 
Here, we exploit the fact that missing data in a continuous covariate is an extreme case of classical ME, allowing us to use existing methodology that accounts for ME via a Bayesian framework that employs integrated nested Laplace approximations (INLA), and thus to simultaneously account for both ME and missing data in the same covariate. As a useful by-product, we present an approach to handle missing data in INLA, since this corresponds to the special case when no ME is present. 
In addition, we show how to account for Berkson ME in the same framework. In its broadest generality, the proposed joint Bayesian framework can thus account for Berkson ME, classical ME, and missing data, or for any combination of these in the same or different continuous covariates of the family of regression models that are feasible with INLA. 
The approach is exemplified using both simulated and real data. We provide extensive and fully reproducible Supplementary Material with thoroughly documented examples using \texttt{R-INLA} and \texttt{inlabru}.

\end{abstract}


\section{Introduction}\label{intro}

Missing data and measurement error (ME) are both cases of information loss. This loss means that at the very least, the level of uncertainty in our inferences increases, and, depending on the underlying mechanism, the error and missing data might also be biasing our conclusions in ways that are not immediately obvious \citep{fuller1987, little_rubin1987, carroll_etal2006}. Both problems are unavoidable in any field that deals with observational data. A common example is missing responses in surveys, where ME may also be present due to imprecise or untrue answers. Other examples can be found in exposure assessment in epidemiology \citep{heid_etal2004, goldman_etal2011}, phenotypic ME in quantitative genetics \citep{ponzi_etal2018}, or errors in expert coded data in political science \citep{marquardt2020}. It is therefore important to understand how to evaluate the impact of ME and missing data on statistical inference, as well as how to adjust for both to ensure valid inference. 

Missing data and its implications for statistical analyses have been studied extensively \citep[\eg,][]{little_rubin1987, vanbuuren2018, carpenter_smuk_2021}.
The missingness of values in a covariate can be classified based on three main mechanisms that generate the missingness. The first type is referred to as \emph{missing completely at random} (MCAR), arising when the missingness occurs entirely at random, not depending on any variables, observed or unobserved. The second type, referred to as \emph{missing at random} (MAR), occurs when the missingness depends on some other variable(s) which are themselves entirely observed. The last type is referred to as \emph{missing not at random} (MNAR), and occurs when the missingness depends on the values of the variables that are themselves unobserved \citep{little_rubin1987}. 
In a separate branch of research, ME has also been studied extensively \citep[\eg,][]{fuller1987, gustafson2003, carroll_etal2006, buonaccorsi2010, yi2017}. Apart from increasing uncertainty, ME typically introduces bias into parameter estimates, whereas the type and direction of bias heavily depends on the actual error mechanism. As a major distinction, we can classify ME into classical and Berkson error, which influence parameter estimates in fundamentally different ways and must therefore be treated specifically \citep[see, \eg,][]{berkson1950, carroll_etal2006}. There may be cases where the ME does not influence the conclusions, but this cannot be known without inspection. It is therefore recommended to model the ME or carry out a simulation study when one suspects ME in the data in order to get more insight into the effect on estimates and their uncertainty \citep{vansmeden_etal2019, stratos1_2020, innes_etal2021}.

When dealing with ME or missing data, Bayesian methods enable us to specifiy models that mirror the data generating process, and we can easily include models that explain these errors in the data jointly along with the actual models we are interested in.
An imputation model, for example, can be fit as part of a hierarchical Bayesian model \citep{erler_etal2016, ma_chen_2018}, and thus propagates the added uncertainty appropriately through the levels. Similarly, if there is ME, we may specify an error model describing the relation between the correct variable and the observed one, and incorporate any information we have about the ME through priors \citep{richardson_gilks1993, dellaportas_stephens1995, muff_etal2015}.

While most of the literature on missing data and ME does not overlap, the similarities between the two problems have recently been pointed out. As an example, \citet{cole_etal2006} present multiple imputation for ME correction. In another step forward, \citet{blackwell_etal2017} treat ME and missing data in the same framework by using the observation that missing data is a limiting case of ME, corresponding to the case when the error variance tends to infinity, that is, the complete absence of information (\emph{i.e.}, missingness).  
Soon after, \citet{goldstein_etal2018} described a hierarchical Bayesian model with a classical ME and imputation model that allows the covariate with missing values to depend on other variables, and \citet{keogh_bartlett2019} considered ME as a missing data problem by demonstrating multiple ways to utilize this link. \citet{noghrehchi_etal2020} combined multiple imputation with functional methods for error correction in a joint framework, and most recently \citet{vansmeden2021} compared four different methods that address both ME and missing data, among them a Bayesian model. 
The integrated nested Laplace approximation framework enables Bayesian inference for complex hierarchical models and has over the past decade become a popular tool for a wide variety of statistical models \citep{rue_etal2009, martino_riebler2020, gomezrubio2020}. 
In the context of INLA, both missing data and ME in covariates are a challenge, as the covariates are part of the latent Gaussian Markov random field (GMRF), and INLA requires the values in this latent field to be fully observed \citep{gomezrubio_etal2019}. No joint treatment of missing data and ME in INLA has previously been proposed, although methodology for missing data and ME has been suggested separately. In \citet{berild_etal2022}, importance sampling is combined with INLA, which then allows for missing data imputation. However, in scenarios where many observations are missing this becomes unfeasible. On the other hand, \citet{gomezrubio_etal2019} present a general approach by defining a covariate imputation model as a latent effect with a GMRF structure. 
In the case of covariate ME, \citep{muff_etal2015} show how classical and Berkson ME in continuous covariates can be incorporated into the INLA framework.

Here, we propose to take advantage of the ME and missing data connection to provide joint ME and missing data modeling using INLA. 
To this end, we use the interpretation of missingness as an extreme case of classical ME in order to employ existing ME models in INLA for missing data imputation.
In terms of missing data, our approach is less general than the one by \citet{gomezrubio_etal2019}, since their approach also allows for modelling of the missingness mechanism, which is necessary when the missingness is MNAR. However, in many practical applications it is reasonable to make the assumption that the missingness is MAR or even MCAR, in which case the missingness mechanism is ignorable. The model presented in this paper is therefore well suited for those cases, and has the advantage of being able to account for ME simultaneously. The proposed Bayesian model can be fit in any framework for Bayesian modeling, but we especially highlight the necessary adjustments that must be made in INLA.
This paper is organized as follows. In Section \ref{model_specification}, we describe the Bayesian model that will be used to model ME and missing data in covariates. In Section \ref{sec:fitting_inla} we explain how the model is implemented in INLA. In Section \ref{applications}, we illustrate the model and its implementation with three examples. Lastly, Section \ref{discussion} provides a discussion of the method, results, limitations and extensions.

\section{Model specification}\label{model_specification}

\subsection{Classical measurement error}\label{sec:classic}

In the case of classical ME, we are interested in $n$ observations of a variable $\bm{x} =  (x_1, x_2, \dots, x_n)^T$, but we actually observe (in vector notation) $\bm{w} = \bm{x} + \bm{u}_c$, with an additive noise term $\bm{u}_c \sim \mathcal{N}(\bm{0}, \tau_{u_c}\bm{D}_{u_c})$ that is independent of $\bm{x}$. The subscript $c$ is used here to indicate that it is the error term in a \emph{classical} error model, $\tau_{u_c}=1/\sigma_{u_c}^2$ denotes the precision of the error term, and $\bm{D}_{u_c}$ may capture heteroscedasticity or dependence structures, but in the simplest case $\bm{D}_{u_c}$ is the identity matrix, implying constant ME precision across all observations. A characteristic of classical ME is that the variance of the observed variable $\bm{w}$ can be additively split into $\sigma^2_w = \sigma_x^2 + \sigma_{u_c}^2$, thus the variance of $\bm{x}$ will be smaller than the variance of the misobserved version $\bm{w}$. 

When $\bm{x}$ is a covariate in a regression model, a Bayesian hierarchical model to account for classical ME in the variable $\bm{x}$ can be formulated as \citep[see \eg,][]{richardson_gilks1993, dellaportas_stephens1995, muff_etal2015}
\begin{align}
  \bm{\eta} &= \beta_0 \bm{1} + \beta_x \bm{x} + \bm{Z} \bm{\beta}_z \ , & \label{eq:regmodc} \\
  \bm{w} &= \bm{x} + \bm{u}_c \ , &\bm{u}_c \sim \mathcal{N}(\bm{0}, \tau_{u_c}\bm{D}_{u_c}) \ , \label{eq:errormodc} \\
  \bm{x} &= \alpha_0 \bm{1} + \widetilde{\bm{Z}}\bm\alpha_z + \bm{\varepsilon}_x \ , &\bm{\varepsilon}_x \sim \mathcal{N}(\bm{0}, \tau_{\varepsilon_x}\bm{D}_{\varepsilon_x}) \ , \label{eq:expmodc}
\end{align}
where the components are: 
\begin{description}
  \item[Regression model \eqref{eq:regmodc} :]{ 
  $g(\bm{\mu}) = \bm{\eta}$ is the linear predictor in a generalized linear model (GLM), given the correct covariate values for $\bm{x}$, as well as other covariates observed without error, stored in the matrix $\bm{Z}$. Here, $g(\cdot)$ is the link function and $\bm{\mu}$ is the mean vector. The coefficients $\beta_0$, $\beta_x$ and $\bm\beta_z$ are the intercept and the slopes for $\bm{x}$ and $\bm{Z}$, respectively, and these may be given Gaussian priors centered at $0$ with small precisions. 
  }
  \item[Classical error model \eqref{eq:errormodc} :]{The error model describes how the covariate with ME is related to the correct variable, where $\bm{u}_c$ is the error in the observed variable $\bm{w}$, and $\bm{x}$ and $\bm{u}_c$ are assumed to be independent. Importantly, the prior of the precision $\tau_{u_c}$, typically a gamma prior, must be chosen based on expert knowledge, validation data or repeated measurements to ensure identifiability \citep{gustafson2005, gustafson2010}.
  }
  \item[Imputation model \eqref{eq:expmodc} :]{Describes how the covariate with error might depend on other fully observed covariates, stored in the matrix $\widetilde{\bm{Z}}$ (which may or may not be equal to $\bm{Z}$). Here, $\bm\varepsilon_x$ is the residual term in the model for the covariate $\bm{x}$, and its precision $\tau_{\varepsilon_x}$ should be given an carefully chosen prior. The intercept $\alpha_0$ and slopes $\bm\alpha_z$ for the covariates in $\widetilde{\bm{Z}}$ may be given Gaussian priors centered at $0$ with small precisions. In the ME literature, Model \eqref{eq:expmodc} is generally referred to as the \emph{exposure} model.
  }
\end{description}

\subsection{Missing data in relation to classical measurement error}\label{sec:missing}

\begin{figure}[h]
  \centering
  \includegraphics[width=10cm]{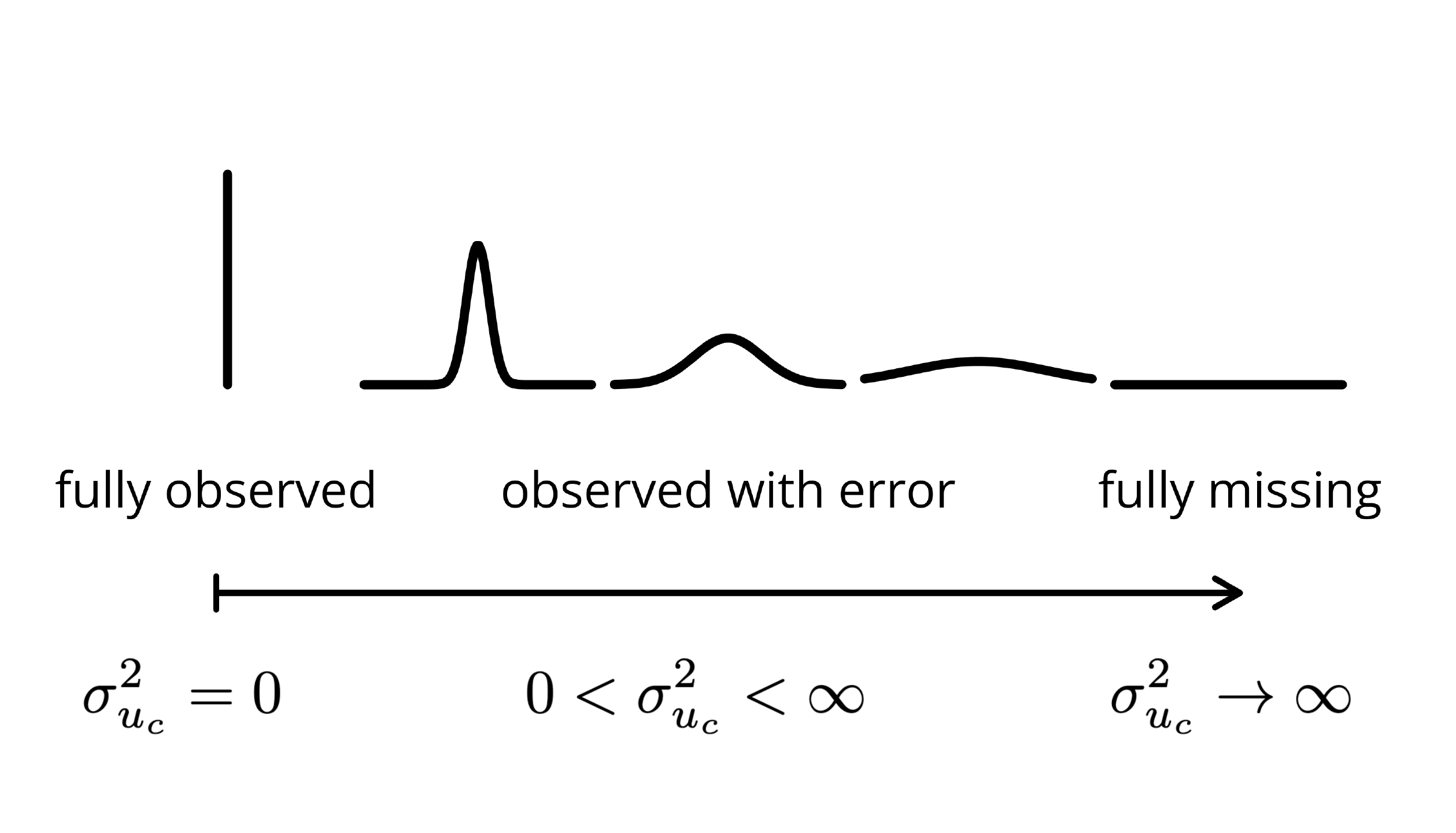}
  \caption{The scale of classical measurement error variance. The figure is adapted with permission from \citet{blackwell_etal2017} (copyright 2015 by the authors).}  
  \label{fig.scaleme}
\end{figure}

A convenient feature of the ME model in Section \ref{sec:classic} is that it can at the same time be used to account for missing values. Imagine that the classical ME variance $\sigma_{u_c}^2$ exists on a continuum where we have an observation with no error when the variance is zero ($w_i = x_i$), and as the variance increases our ME gradually increases, to the point where we have no information about the observation, which is analogous to missing data (Figure \ref{fig.scaleme}). We can thus think of missing data as an extreme case of ME. Using this, we can reflect the degree of error through the prior of the classical error variance: either there is no error ($\sigma_{u_c}^2 = 0$), some finite error ($0<\sigma_{u_c}^2<\infty$), or an observation is completely missing ($\sigma_{u_c}^2 \rightarrow \infty$).  We cover the technical aspects of this in Section \ref{sec:scaling_inla}. Note that Figure \ref{fig.scaleme} shows the ME variance $\sigma_{u_c}^2$ for illustration purposes, whereas we use the precision $\tau_{u_c} = 1/\sigma_{u_c}^2$ in the rest of the paper. 

\subsection{Berkson measurement error}\label{sec:berkson}
The Berkson ME \citep{berkson1950} is fundamentally different from the classical ME. While the classical ME can be thought of as random noise, the Berkson ME arises for example in situations where all members of a population are assigned the same value for some property, when in reality the value varies between the members. Berkson error is quite typical in experimental setups, where a variable is intended to be fixed at a given value, such as a concentration or a specific time point, when in reality there is some level of variation due to practical deviations from the target value \citep[see \eg,][]{buonaccorsi_2002}. 
The Berkson error model can be formulated as $\bm{x} = \bm{w} + \bm{u}_b$, where $\bm{x}$ is as before the correct value, $\bm{w}$ is the observed value and $\bm{u}_b \sim \mathcal{N}(\bm{0}, \tau_{u_b}\bm{D}_{u_b})$, with Berkson error precision $\tau_{u_b}=1/\sigma_{u_b}^2$ that scales the precision matrix $\bm{D}_{u_b}$. Assuming $\bm{w}$ and $\bm{u}_b$ to be independent, the variance of the correct obervations is additively split like $\sigma_x^2=\sigma_w^2 + \sigma_{u_b}^2$. A property of the Berkson ME is thus that the variance of the observed values $\bm{w}$ will be smaller than the variance of the correct value $\bm{x}$, as opposed to the classical error situation. As with the classical ME, we can set up a Bayesian hierarchical model that can be used to account for Berkson ME in the variable $\bm{x}$ \citep{muff_etal2015}, 
\begin{align}
  \bm{\eta} &= \beta_0 \bm{1} + \beta_x \bm{x} + \bm{Z} \bm{\beta}_z \ , & \label{eq:regmodb} \\
  \bm{x} &= \bm{w} + \bm{u}_b \ , &\bm{u}_b \sim \mathcal{N}(\bm{0}, \tau_{u_b}\bm{D}_{u_b}) \ , \label{eq:errormodb} 
\end{align}
where Model \eqref{eq:regmodb} is the regression model of interest (and corresponds to Model \eqref{eq:regmodc} above), and Model \eqref{eq:errormodb} is the Berkson ME model with independent error term $\bm{u}_b$. Note that we do not need an imputation model, as $\bm{x}$ is defined conditionally on the observations $\bm{w}$ in the error model \eqref{eq:errormodb}.

\subsection{A generalized measurement error model}\label{sec:combined}
In some cases both error types occur together, for instance in exposure assessment in epidemiology \citep[see \eg,][]{reeves_etal1998, heid_etal2004, deffner_etal2018}. By combining the hierarchical classical error model given by components \eqref{eq:regmodc} -- \eqref{eq:expmodc} with the Berkson ME model \eqref{eq:regmodb} -- \eqref{eq:errormodb} it is possible to handle classical and Berkson ME in the same covariate. We again denote the correct variable as $\bm{x}$ and the observed version by $\bm{w}$, but we need to introduce an additional latent variable $\bm{r}$ \citep{muff_etal2017}, where the relation between $\bm{r}$ and the correct covariate $\bm{x}$ is the Berkson error model $\bm{x} = \bm{r} + \bm{u}_b$. This leads to the combined model formulated as 
\begin{align}
  \bm{\eta} &= \beta_0 \bm{1} + \beta_x \bm{x} + \bm{Z} \bm{\beta}_z \,, & \label{eq:regmod} \\
  \bm{x} &= \bm{r} + \bm{u}_b \,, & \bm{u}_b \sim \mathcal{N}(\bm{0}, \tau_{u_b}\bm{D}_{u_b}) \ , \label{eq:berrormod}  \\
  \bm{w} &= \bm{r} + \bm{u}_c \,, & \bm{u}_c \sim \mathcal{N}(\bm{0}, \tau_{u_c}\bm{D}_{u_c}) \ , \label{eq:cerrormod}  \\
  \bm{r} &= \alpha_0 \bm{1} + \widetilde{\bm{Z}}\bm{\alpha}_z + \bm{\varepsilon}_x \,, &\bm{\varepsilon}_r \sim \mathcal{N}(\bm{0}, \tau_{\varepsilon_r}\bm{D}_{\varepsilon_r}) \ . \label{eq:expmod}
\end{align}

Note that the imputation model is defined for $\bm{r}$, as this is the latent variable entering the classical ME model, whereas the model for the latent "correct" variable $\bm{x}$ is given through the Berkson ME model. To better understand the role of the latent variable $\bm{r}$, it may help to imagine what would happen if one of the error terms was "turned off" in the model. If we have $\bm{u}_b = \bm{0}$, then \eqref{eq:berrormod} simply says $\bm{x} = \bm{r}$, and we are left with the classical error model we introduced in Section \ref{sec:classic}. On the other hand, if there is no classical error ($\bm{u}_c = \bm{0}$) then $\bm{w} = \bm{r}$, and we are only left with the Berkson error model. In that case, the coefficients of the imputation model will be estimated, but it will not feed back into the model, so in practice it does not affect $\bm{x}$. If we have Berkson error and missing data, we need to keep the classical error model in order to to impute the missing values, but we scale the classical ME precision to a large value for the values that are observed to indicate that there is no classical ME. The handling of the missingness is then done just as described in Section \ref{sec:missing}. 

\subsection{Possible extensions}
The model described in Section \ref{sec:combined} can in the most general case account for classical ME, Berkson ME, and observations missing at random in the same variable, with the regression model of interest being any generalized linear model that is compatible with the assumptions INLA takes. There are some straightforward extensions to this -- if there is ME or missingness in multiple covariates, then separate error and imputation models need to be formulated for each such covariate. If the error in one covariate is heteroscedastic, then the elements of $\bm{D}_{u_c}$ can correspondingly be adjusted to reflect the differing precisions for the different observations. Biased errors can be specified in the error model by introducing an offset. 
Due to the modular structure of Bayesian hierarchical models, the components may be combined with any other model formulations that are possible within a Bayesian framework, for example random effects or spatial effects \citep[see \eg,][]{gryparis_etal2008}.

\section{Technical considerations for missing and mismeasured covariates in INLA} \label{sec:fitting_inla}


A requirement to use INLA is that the model can be expressed as a latent Gaussian Markov random field, which makes INLA slightly more restrictive than standard sampling approaches. There are also some practical issues that arise when fitting the particular model described in Section \ref{sec:combined}, which we will now address.

\subsection{How are missing values treated in INLA?}\label{sec:missingINLA}
The technical issue of how missing values are handled in INLA is not addressed in the original INLA publication \citep{rue_etal2009}, but the details were later discussed, for example on the FAQ website \url{https://www.r-inla.org/faq}, question 7: "How does inla() deal with 'NA's in the data-argument?". 
First of all, it is crucial to discriminate between missing values in the response versus missingness in the covariates. If a value is missing in the response, then that data point gives no contribution to the likelihood, while the predictive distributions for these missing values in the response will be computed automatically \citep[][Chapter 12.3]{gomezrubio2020}. If a value in a covariate is missing, on the other hand, that entry is re-coded to be $0$ internally. This way of handling missing values in the covariates may lead to biased posteriors, and it is the case we address here.

As mentioned in the introduction, a few publications have tackled missing values in covariates in INLA previously. \citet{gomezrubio_rue2018} combine Markov chain Monte Carlo with INLA to this end, and \citet{berild_etal2022} show how to combine importance sampling with INLA, and both use missing covariate imputation to illustrate their approaches. However, in both cases the missing observations are treated as additional parameters of the model, and thus the model becomes very computationally expensive when the number of missing observations is large, which makes it of limited practical use. The method proposed in \citet{gomezrubio_etal2019}, on the other hand, is based purely on INLA, and is much more widely applicable. They also include a logistic model to describe the actual missingness mechanism, which means that it is possible to do a sensitivity analysis to examine the influence of the missingness. This is advisable when one suspects that the missingness mechanism may be MNAR, that is, when the variable is missing depending on the value of the variable itself. However, in the cases where we have reason to believe that missingness is MCAR or MAR it is not necessary to include a model for the missingness mechanism itself \citep{vanbuuren2018}, and we can employ a simpler model than \citet{gomezrubio_etal2019}. In that case, we can take advantage of the interpretation of missing data as an extreme form of ME used here.

\subsection{Scaling the classical error precision for the case of missing values}\label{sec:scaling_inla}
In Section \ref{sec:missing} we described the interpretation of missingness as an extreme form of ME. Conceptually, scaling the classical error precision to a large value (\eg, $10^{12}$) would indicate that we have a small error, while scaling to a very small value (\eg, $10^{-12}$) would correspond to a missing observation, since we have very large uncertainty. Values for the precision in between these two extremes would indicate a regular classical ME. The respective information can thus directly be incorporated into the corresponding element of the (scaled) diagonal matrix $\tau_{u_c}\bm{D}_{u_c}$. However, in practice it is even not necessary to set the precision to a small value if a covariate entry is missing, because the values that are completely missing in $\bm{w}$ do not contribute to the likelihood in the error model component \eqref{eq:cerrormod} thanks to the way missing values are handled in INLA (see Section \ref{sec:missingINLA}). We thus only need to scale the precision according to whether a given observation has ME or not. In the case when the value is missing, the imputation model will impute the missing value for $w_i$ based on the other covariates through the imputation and error models in \eqref{eq:expmod} and \eqref{eq:cerrormod}.

\subsection{Reformulating sub-models that have latent variables as responses}
In both missing data and ME problems, the actual covariate $\bm{x}$ is unobserved and therefore part of the latent field. In a standard Bayesian hierarchical model, the imputation model imputes the unobserved values based on other available covariates. 
However, imputation models like the one presented in Equation \eqref{eq:expmod} cannot be formulated directly in INLA, since the framework does not allow latent variables in the response.
For the case of ME, a solution to this problem has however already been suggested by \citet{muff_etal2015}: the imputation model in Equation \eqref{eq:expmod} is reformulated with pseudo-observations $\bm{0}$ as the response, which circumvents the problem as the imputation model can now be handled as part of the observation model. The same workaround needs to be used for the specification of a Berkson ME model, where the unobserved variable $\bm{x}$ would otherwise stand as the response. Our full model as formulated in INLA then becomes
\begin{align}
  \bm{\eta} &= \beta_0 \bm{1} + \beta_x \bm{x} + \bm{Z} \bm{\beta}_z \ , & \\
  \bm{0} &= - \bm{x} + \bm{r} + \bm{u}_b \,, & \bm{u}_b \sim \mathcal{N}(\bm{0}, \tau_{u_b}\bm{D}_{u_b}) \ , \\
  \bm{w} &= \bm{r} + \bm{u}_c \,, & \bm{u}_c \sim \mathcal{N}(\bm{0}, \tau_{u_c}\bm{D}_{u_c}) \ , \\
  \bm{0} &= -\bm{r} + \alpha_0 \bm{1} + \widetilde{\bm{Z}}\bm{\alpha}_z + \bm{\varepsilon}_x \ , \label{eq:trick} &\bm{\varepsilon}_x \sim \mathcal{N}(\bm{0}, \tau_{\varepsilon_x}\bm{D}_{\varepsilon_x}) \ .
\end{align}
The same approach is directly applicable to the case with missing data.
Another issue that arises is the estimation of the coefficient $\beta_x$, which is not straightforward since $\bm{x}$ is unobserved. In fact, although both $\bm{x}$ and $\beta_x$ are assumed to be Gaussian distributed, their product is not, while INLA requires that the latent field is Gaussian. Again, a solution to this was presented in \citet{muff_etal2015} for the ME case, and thus the approach is directly applicable to the case with missing data as well. In brief, instead of estimating $\beta_x$ as a part of the latent field, we let it be a hyperparameter, which ensures that the latent field is still Gaussian, conditioned on $\beta_x$.

\section{Applications}\label{applications}

In this section we examine two real-world data sets and carry out a simulation study. Code for all three examples is available in the Supplementary Material, where each example is presented in separate documents with extensive explanations.

\subsection{Missing covariate imputation: a linear model for cholesterol}\label{sec:bmi}
In order to illustrate how to model missing covariate imputation in INLA using the framework of a classical ME model, we consider the common case where only missing data is present, but no ME.
We consider the \verb|nhanes2| (National Health and Nutrition Examination Survey) dataset from the \verb|mice| R package \citep{vanbuuren_etal2011}, which has often been used to evaluate missing data imputation methods, \citep[see, \eg,][]{vanbuuren_etal2011, gomezrubio_rue2018, gomezrubio_etal2019, berild_etal2022}. 
The data contains 25 observations and four variables: age group (which is categorical with three levels), BMI, hypertension and cholesterol. Note that this data set is not large enough to make inference. The example is provided as an illustration for how to implement the method for missing covariate data, since the data set is commonly used for this purpose. Here, we fit a linear regression model with cholesterol (chl) as the response, and age group and BMI as covariates. The response cholesterol and the covariate BMI have 10 and 9 missing observations, respectively, but since the missingness in the response is automatically handled by INLA (see Section \ref{sec:missingINLA}), the problem reduces to imputing the missing values in the BMI variable. 
We specify the joint Bayesian model with a regression model for cholesterol \eqref{ex.missing.moi}, the error model \eqref{eq:error_chol} and the imputation model \eqref{ex.missing.imputation}, where the latter describes how BMI depends on age category:
\begin{align}
\label{ex.missing.moi}
\mathit{chl}_i &= \beta_0 + \beta_1 \mathit{age}^{(40-59)}_i + \beta_2 \mathit{age}^{(60-99)}_i + \beta_3 \mathit{bmi}_i + \varepsilon_i  \ , \quad &\varepsilon_i \sim \mathcal{N}(0,\tau_y) \ , \\
\label{eq:error_chol}
\mathit{bmi}_i^{obs} &= \mathit{bmi}_i + u_i \ , \quad & u_i \sim \mathcal{N}(0, \tau_u \bm{D}_{ii}) \ , \\
\label{ex.missing.imputation}
 \mathit{bmi}_i &= \alpha_0 + \alpha_1 \mathit{age}^{(40-59)}_i + \alpha_2 \mathit{age}^{(60-99)}_i + \varepsilon^{(x)}_i \ , \quad &\varepsilon^{(x)}_i \sim \mathcal{N}(0,\tau_x) \ .
\end{align}
Here, $\mathit{bmi}^{obs}_1, \dots \mathit{bmi}^{obs}_n$ are our observations, which contain missing values, whereas $\mathit{bmi}_1, \dots \mathit{bmi}_n$ are the latent variables. Age is dummy-coded for the three age-groups. 
The coefficients in Model \eqref{ex.missing.moi} and Model \eqref{ex.missing.imputation} are all given $\mathcal{N}(0, 10^{-6})$ priors. The priors for the precisions of the residuals $\bm{\varepsilon}$ and $\bm{\varepsilon}^{(x)}$ are 
$\tau_y \sim \Ga(2, 846.8)$ and $\tau_x \sim \Ga(2, 16.8)$, chosen based on information from the data (this is described in detail in the Supplementary Material), while the precision for the error term, $\tau_u$, is fixed to 1. Since we have no ME in this example, and since the missing values for $\mathit{bmi}^{obs}_i$ (which is the response of the ME model \eqref{eq:error_chol}) do not contribute to the likelihood, we can actually set \emph{all} the diagonal elements of $\bm{D}$ to $10^{12}$.

\begin{figure}[t]
  \centering
  \includegraphics[width=\textwidth]{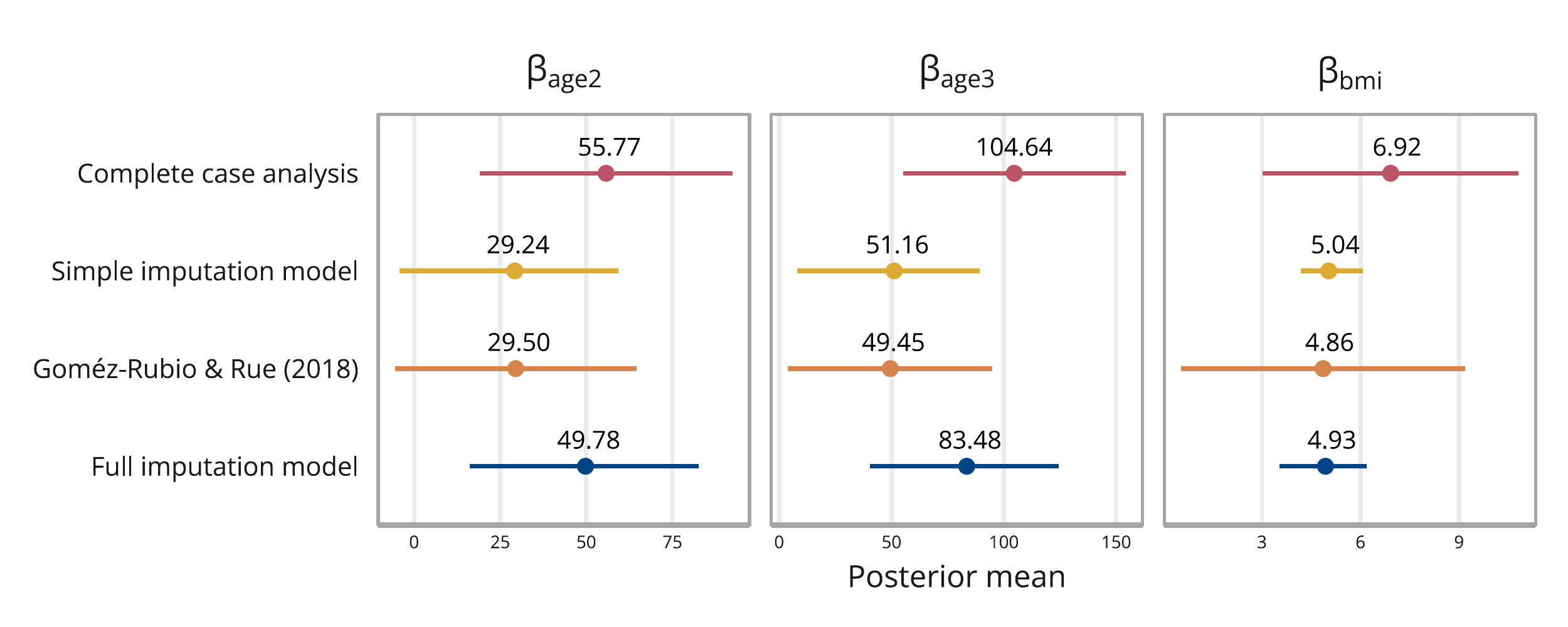}
  \caption{Posterior means and 95\% credible intervals for the parameters in Section \ref{sec:bmi}. The complete case analysis (red) is a model fit using only the complete cases ($n=16$). The simple imputation model (yellow) uses a Gaussian prior for the missing observations, and is identical to the model used in \cite{gomezrubio_rue2018} (orange) ($n = 25$). The full imputation model (blue) also depends on the age of the patient for the imputation of the missing BMI values ($n = 25$).}
  \label{fig:missing}
\end{figure}

For comparison, we fit two additional models, one complete case model that uses only the observed BMI values, as well as a simpler imputation model, where the imputation is independent of age (Figure \ref{fig:missing}). In this simpler model, 
the missing body mass values have a simple Gaussian prior centered at the average of the observed values and variance four times the variance of the observed values, as in \citet{gomezrubio_rue2018}. For completeness, we also report the exact results from \citet{gomezrubio_rue2018}. Across all coefficients, the posterior means for the complete case model are consistently larger than when an imputation model is included. It is to be expected that uncertainties decrease when we are able to include more observations, however the very small credible intervals in this case are more likely an inconsistency due to the very small sample size. Overall, the example illustrates that our approach implemented in INLA is a valid alternative to existing tools for missing data imputation.

\subsection{Imprecise and missing systolic blood pressure measurements}\label{sec:bloodpressure}

We now consider a case of ME and missing values in the covariates of a survival model, using the framework described in Section \ref{model_specification}.  The data is originally from the Third National Health and Nutrition Examination Survey (NHANES III), but we use the data as pre-processed and provided by \citet{bartlett_keogh2018}. They linked the NHANES III data to data from the US National Death Index, with information about the mortality status of each participant in 2011. Following \citet{keogh_bartlett2019}, we consider a Weibull survival model with the continuous covariates systolic blood pressure (SBP)  
and age, and binary covariates diabetes status, sex and smoking status. The response is time until death by cardiovascular disease. Deaths by other causes are treated as censorings. 
Measurements of SBP are known to vary substantially within the same patient, and SBP is therefore measured twice for some participants, enabling us to estimate the variance of the error. However, for some participants there is only one measurement, and for others the SBP is completely missing. Note that the smoking status is also missing for around half of the participants, but since we only consider continuous ME, all observations with missingness in the (binary) smoking status are removed for this illustrative example, resulting in $n = 3433$ observations. 



To model the survival time we use a Weibull survival model with hazard $h(t_i) = rt^{r-1}\lambda_i^r$, where $r$ is the shape parameter of the Weibull distribution and $t_i$ is the time until death by cardiovascular disease for individual $i$. The linear predictor $\eta_i$ for individual $i$ is linked to the rate parameter by $\log(\lambda_i) = \eta_i$. The joint model including the ME model for SBP is then
\begin{align}
\log(\lambda_i) &= \beta_0 + \beta_{sbp} \mathit{SBP}_{i} + \beta_{sex} \mathit{sex}_i + \beta_{age} \mathit{age}_i + \beta_{smoke} \mathit{smoke}_i + \beta_{diabetes} \mathit{diabetes}_i  \, & \\
\mathit{SBP}^{(1)}_{i} & = \mathit{SBP}_i + u^{(1)}_{i}\,, \\
\mathit{SBP}^{(2)}_{i} & = \mathit{SBP}_i + u^{(2)}_{i}\,, \\
\mathit{SBP}_i &= \alpha_0 + \alpha_{sex} \mathit{sex}_i + \alpha_{age} \mathit{age}_i + \alpha_{smoke} \mathit{smoke}_i + \alpha_{diabetes} \mathit{diabetes}_i + \varepsilon^{(x)}_i \,, 
\end{align}
for $i=1,\ldots , n$, where $\bm{SBP}^{(1)} = (\mathit{SBP}^{(1)}_{1}, \dots, \mathit{SBP}^{(1)}_{n})^T$ and $\bm{ SBP}^{(2)} = (\mathit{SBP}^{(2)}_{1}, \dots, \mathit{SBP}^{(2)}_{n})^T$ are the first and second blood pressure measurements, respectively, with $\bm{SBP}$ being the latent error-free version of the blood pressure, and the remaining four variables that are assumed to be without error: sex, age, smoking status and diabetes status. The ME terms $\bm{u}^{(1)}$ and $\bm{u}^{(2)}$ and the residual term for the imputation model are assumed to be independently distributed as $N(\bm{0}, \tau_u \bm{I})$ and $\bm\varepsilon^{(x)} \sim \mathcal{N}(\bm{0}, \tau_x\bm{I})$, respectively. 
For the analysis using INLA to be comparable to that of \citet{keogh_bartlett2019}, we used the same priors: the coefficients of the model of interest (the betas) are given $\mathcal{N}(0, 10^{-6})$ priors, while the coefficients of the imputation model (the alphas) are given $\mathcal{N}(0, 10^{-4})$ priors, $\tau_u \sim \Ga(0.5, 0.5)$, and $\tau_x \sim \Ga(0.5, 0.5)$. 

\begin{figure}[t]
  \centering
  \includegraphics[width=\textwidth]{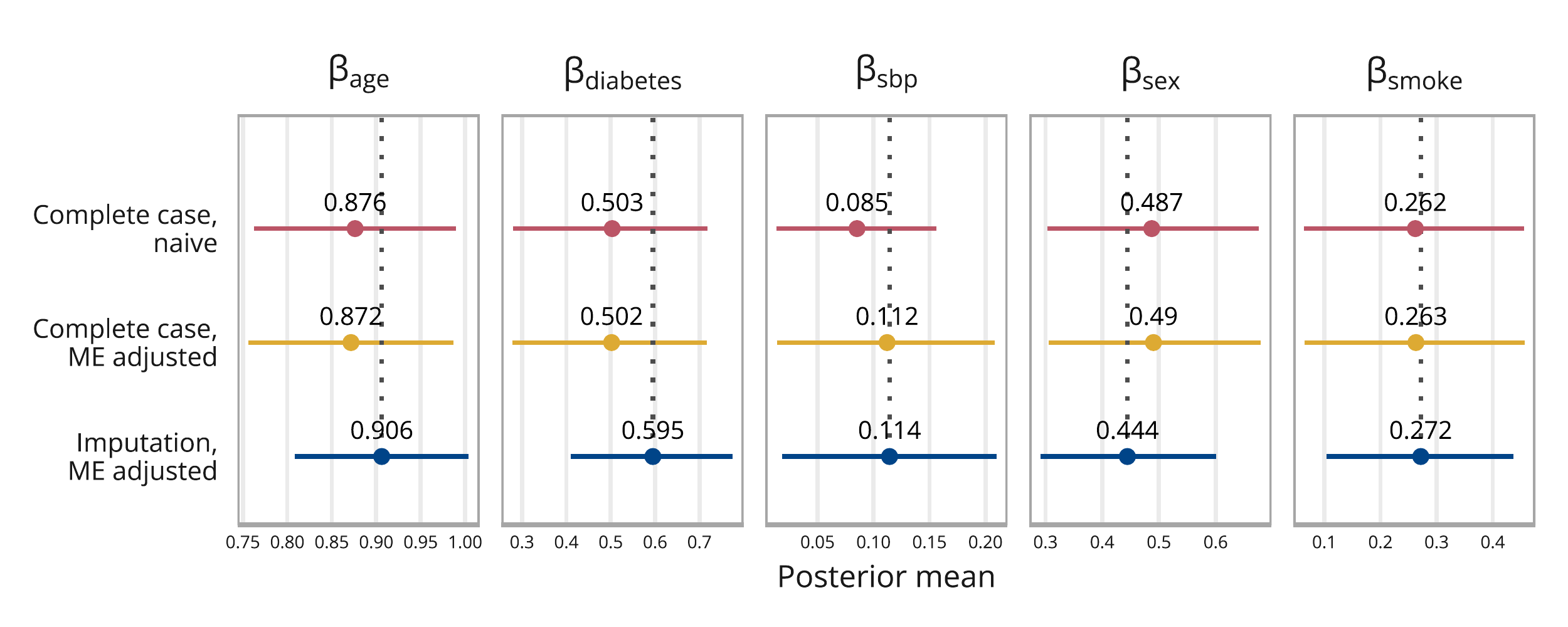}
  \caption{Posterior means and 95\% credible intervals for the parameters in Section \ref{sec:bloodpressure}. The naive complete case analysis (red) is a model fit using only the complete cases ($n = 2667$), and with no ME adjustment. The ME adjusted complete case analysis (yellow) is fit on the same complete case data, but accounting for the ME in the systolic blood pressure. Finally, the ME adjusted and imputation model (blue) uses all the observations ($n = 3433$) and accounts for ME and missing data. A dotted vertical line is drawn from the posterior means of the full imputation and ME model to ease comparison.}
  \label{fig:bloodpressure}
\end{figure}

The model that accounts for ME and missing values ($n = 3433$) is compared to a naive complete case analysis ($n = 2667$) (Figure \ref{fig:bloodpressure}), as well as to the same ME model but with the complete case data, in order to study the effect of accounting for the ME but not the missing observations. 
The naive complete case analysis is only applied to the individuals that had no missing values in $\bm{SBP}^{(1)}$, and treats $\bm{SBP}^{(1)}$ as the correct value for $\bm{SBP}$. We see that the estimated posterior mean for $\beta_{sbp}$ increases in the ME models, which indicates that there may be some attenuation due to the ME. Comparing the ME adjusted complete case model with the model that also imputes the missing values, the coefficients for age and diabetes increase a bit, while the one for SBP shows only a very small increase. It is notable that in this case the missingness seems to be influencing other coefficient estimates than the coefficient of the covariate that actually has the missingness. This may often be the case for both ME and missing data, when the covariate with the ME and missingness is correlated with other covariates. Finally, we compared the runtime of the model fit by \citet{keogh_bartlett2019} in \texttt{JAGS} to the runtime of the model implemented in \texttt{R-INLA}. The model fit in \texttt{JAGS} took 31 minutes (10000 iterations were used, and the first 5000 were discarded), while INLA took 20 seconds (both were run on a 2 GHz Quad-Core Intel Core i5 processor). This improvement shows the significant practical advantage of being able to fit these complex models in \texttt{R-INLA}.

\subsection{Simulation study: Berkson and classical measurement error alongside missing data}\label{sec:simulations}

For this example, we simulated a linear regression model with a mismeasured covariate $\bm{x}$, observed as $\bm{w}$, as well as an error-free covariate $\bm{z}$. For the simulation, we first generated the error free covariate as
\begin{equation}
 \bm{z} \sim \mathcal{N}(\bm{0}, \bm{I}) \ .
\end{equation}
Keeping with the notation used in Section \ref{sec:combined}, the latent variable with Berkson error was then generated as 
\begin{equation}
 \bm{r} =  \bm{1}+2\bm{z} + \bm{\varepsilon}_r \ , \qquad \bm{\varepsilon}_r \sim \mathcal{N}(\bm{0}, \bm{I}) \ ,
\end{equation}
and the correct value for the variable was generated by adding further variation to $\bm{r}$ by
\begin{equation}
 \bm{x} = \bm{r} + \bm{u}_b \ , \qquad \bm{u}_b \sim \mathcal{N}(\bm{0}, \bm{I}) \ .
\end{equation}
The classical ME was then added to the variable with Berkson error to obtain the variable that is actually observed, $\bm{w}$: 
\begin{equation}
 \bm{w} = \bm{r} + \bm{u}_c \ , \qquad \bm{u}_c \sim \mathcal{N}(\bm{0}, \bm{I}) \ .
\end{equation}
Next, the probability for an observation to be missing was assumed to be MAR, depending only on $\bm{z}$ as
\begin{equation}
  \text{P}(w_i = \na) = \frac{\exp(-1.5 + 0.5z_i)}{1+\exp(-1.5 + 0.5z_i)} \ ,
\end{equation}
and the response $\bm{y}$ is generated from the correct covariate $\bm{x}$ as
\begin{equation}
 \bm{y} = \beta_0 \bm{1} + \beta_x \bm{x} + \beta_z \bm{z} + \bm{\varepsilon} \ , \quad \bm{\varepsilon} \sim \mathcal{N}(\bm{0}, \tau_y \bm{I}) \ ,
\end{equation}
with $(\beta_0, \beta_x, \beta_z) = (1, 2, 2)$. 
The simulation was repeated 100 times, with $n = 1000$ observations per iteration. For each data set we fit three different models: A complete case model without error correction using $\bm{w}$ as covariate, a model that accounts for both ME (classical and Berkson) and missing data (and thus reflects the data generating process), and a best-case model that is a regression of $\bm{y}$ on $\bm{x}$ and $\bm{z}$, which shows how this basic regression model would perform if we did not have any ME or missing data. As for the other examples, complete code to reproduce this simulation is available in the Supplementary Material. We also provide an implementation using \texttt{inlabru} rather than \texttt{R-INLA}, since some may find the syntax of \texttt{inlabru} more user-friendly. 

\begin{figure}[h]
  \centering
  \includegraphics[width=15cm]{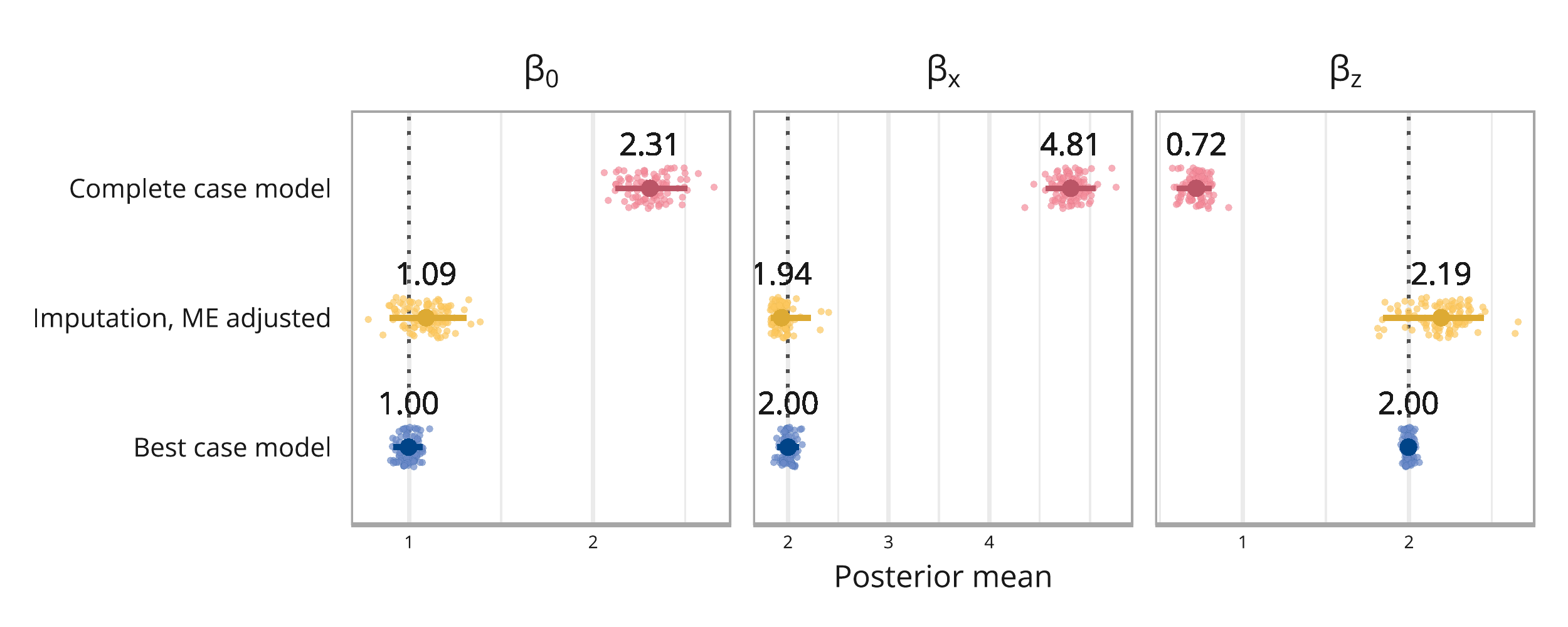}
  \caption{Sample means and 95\% quantile intervals of the posterior means from fitting three different models to 100 different simulated data sets. The dots are the posterior means for each of the runs, and are jittered along the $y$-axis to reduce overplotting. The first model (red) is the naive complete case analysis, simply fitting a regression model using $\bm{w}$ instead of $\bm{x}$, the second model (yellow) adjusts for ME and missing data, and the third model (blue) is the model using the correct value of $\bm{x}$. A dotted vertical line is drawn from the posterior means of the best case model to ease comparison.}
  \label{fig.results.sim}
\end{figure}

The posterior means of the coefficients in the ME model are very close to those actually used for the data generation (Figure \ref{fig.results.sim}), and clearly improved compared to the naive model, for which the posterior mean for $\beta_x$ is over-estimated, while the posterior mean for $\beta_z$ is under-estimated. Although $\beta_z$ is observed without error, the errors in $\bm{x}$ will also greatly bias the estimate for $\beta_z$, since $\bm{z}$ and $\bm{x}$ are correlated. 

\section{Discussion}\label{discussion}

In this paper we have addressed how existing ME modelling approaches for INLA can be used directly to impute missing covariates. By using a model motivated by a ME problem, we are able to work around the practical issue with covariate imputation in INLA. The resulting model formulation provides a way to deal with missingness in a covariate, but it also gives a conceptually simple way to handle cases where ME and missingness both occur in the same variable. In addition to classical ME and missing data, we show how the model can be generalized to account for Berkson ME at the same time. We show that the computational efficiency of INLA allows us to fit the models significantly faster than previously possible in sampling-based software for Bayesian inference.

To reach our goal, we are combining two previously disconnected lines of research in a synergistic way: the first line of research is treating missing data as an extreme case of ME, and the second is doing missing data imputation in INLA. A few publications cited here have used a ME model for missing data imputation in sampling-based inference, but the idea has until now not been incorporated in INLA. Simultaneously, there have been a few different approaches to missing data imputation in INLA, but these have typically not been feasible if the number of missing observations is large. By using the ME model to enable imputation in INLA, we have addressed the computational challenge that was previously a limiting factor.
Moreover, even though we here only considered examples with a single continuous covariate, the model can easily be expanded to accommodate multiple covariates with missing or mismeasured variables. On the other hand, it is not straightforward to account for missingness or ME in categorical covariates, since in INLA the latent variables must be assumed to be Gaussian. Modelling the misclassification and missingness of a categorical variable might be achieved by combining this method with MCMC within INLA.

In summary, we have shown how a joint Bayesian model for ME and missingness allows us to account for both classical and Berkson error in continuous covariates of regression models, as well as providing a new way to do covariate imputation in INLA. A Bayesian analysis gives us the flexibility we need to reflect the error-generating process directly in the model, as well as the option to include prior knowledge about the error variance. Our approach to handle missing data and ME in INLA enables significantly faster inference compared to sampling-based methods, which makes these complex Bayesian models for ME and missing data feasible for a wide range of practical applications.


\section*{Data availability statement}
The data used in this paper is available in the Supplementary Material, along with the \texttt{R}-code to reproduce the results. The Supplementary Material can be found at \url{https://emmaskarstein.github.io/Missing-data-and-measurement-error/}. 

\section*{Acknowledgements}
The authors thank Malgorzata Roos, Ekaterina Poliakova and Christoffer Wold Andersen for insightful comments on an early version of the manuscript.

\section*{Conflict of interest statement}
The authors have declared no conflict of interest.

\newpage

\clearpage

\bibliography{bibliography}

\clearpage

\listoffigures

\end{document}